\documentclass[twocolumn,showpacs,amsmath,amssymb]{revtex4}
\usepackage{graphicx}
\usepackage{dcolumn}
\usepackage{bm}
\newcommand{\beq}{\begin{equation}}
\newcommand{\eeq}{\end{equation}}
\newcommand{\bea}{\begin{eqnarray}}
\newcommand{\eea}{\end{eqnarray}}
\newcommand{\eps}{\varepsilon}

\begin{document}

\author{S. V. Tolokonnikov}
\affiliation{Kurchatov Institute, 123182 Moscow} \affiliation{Moscow
Institute of Physics and Technology, 123098 Moscow, Russia}
\author{S. Kamerdzhiev}
\affiliation{Institute for Physics and Power Engineering, 249033
Obninsk, Russia}
\author{S. Krewald}\affiliation{Institut f\"ur Kernphysik, Forschungszentrum J\"ulich,
D-52425 J\"ulich, Germany}
\author{E. E. Saperstein}
\affiliation{Kurchatov Institute, 123182 Moscow}
\author{D. Voitenkov}
\affiliation{Institute for Physics and Power Engineering, 249033
Obninsk, Russia}

\title{ Quadrupole moments of spherical semi-magic nuclei
within the self-consistent Theory of Finite Fermi Systems}


\begin{abstract}
 The quadrupole moments  of odd neighbors of semi-magic lead and tin
isotopes and $N=50,N=82$ isotones are calculated within the
self-consistent Theory of Finite Fermi Systems based on the Energy
Density Functional by Fayans et al. Two sets of parameters, DF3 and
DF3-a, fixed previously are used. They differ by the spin-orbit and
effective tensor force parameters, the latter being significantly
bigger in the DF3-a functional. Results for the two functionals
turned out to be rather different. The functional DF3-a leads to
quadrupole moments in reasonable agreement with the experimental
ones for most nuclei examined.

\end{abstract}

\maketitle

\section{Introduction}
In the last decade, the interest for
ground state nuclear properties, in particular, static moments, was renewed
due to modern Radioactive Ion Beam facilities which provide
access to long chains of isotopes, including the radioactive ones, in
their ground and isomeric states. Such nuclei distant from the $\beta$-decay
stability valley and often very close to the drip lines are of great
interest for nuclear astrophysics.
Spectroscopy techniques using high
intensity lasers allow for precision measurements of nuclear spins,
magnetic and quadrupole moments which are much more delicate nuclear characteristics
than masses and charge radii. As a result, the bulk of the data on nuclear
static moments becomes very extensive and comprehensive
\cite{Stone} creating a challenge to nuclear theory. Recent precise measurements
of magnetic and quadrupole moments of a long chain of copper isotopes from
$N=28$ to $N=46$ presented and analyzed in the comprehensive article \cite{Q_Cu} is a new evidence of this
interest. For nuclear magnetic moments, this challenge
was partially responded recently \cite{mu1,mu2}. Data on magnetic
moments of more than one hundred odd nuclei were reproduced with an accuracy of
$0.1{-}0.2 \mu_{\rm N}$  within the self-consistent Theory of Finite Fermi Systems (TFFS).
In this study, the QRPA-like TFFS equations for the nuclear response to the external magnetic field
were solved on the base of the Energy Density Functional (EDF) by Fayans et al.
\cite{Fay1,Fay5,Fay}. Especially high accuracy was reached for semi-magic
nuclei considered in the ``single-quasiparticle approximation''
where one quasiparticle in the fixed state $\lambda=(n,l,j,m)$ with
the energy $\varepsilon_{\lambda}$ is added to the even-even core. Such nuclei
where only one kind of nucleons is superfluid are, as a rule, spherical. In addition,
the lowest $2^+$-state has usually an excitation energy of the order 1 MeV, the corresponding
collectivity is not extremely  high and the particle-phonon coupling is not
sufficiently strong to destroy the single-quasiparticle approximation.
 According to the TFFS \cite{AB1},  a quasiparticle  differs  from
 a particle of the single-particle model in two respects.
First, it possesses a local charge $e_q$  and, second, the core is polarized
 due to the interaction between the particle and the core nucleons via the
 Landau--Migdal (LM) amplitude. In other words, the quasiparticle possesses the
 effective charge $e_{\rm eff}$ caused by the polarizability  of the core which is found
 by solving the TFFS equations.
 In the many-particle Shell Model \cite{BABr},
 a similar quantity is  introduced  as a phenomenological parameter which describes
 polarizability of the core consisting of  outside nucleons. For nuclei with both non-magic
  sub-systems corrections to the single-quasiparticle approximation due to
 $2^+$-phonon coupling are important and should be taken into account to
 describe magnetic moments correctly \cite{mu1,mu2}.

For quadrupole moments, the  systematic calculations we know concern
only the medium atomic weight nuclei of the $2p-1f$ shell
\cite{BABr} and $2p,1f_{5/2},1g_{9/2}$ shell \cite{Q_Cu}. Evidently,
the first step in this direction for heavy nuclei was made in our
recent article \cite{BE2} where we analyzed effects of the density
dependence by studying the $2^+_1$-excitations in two isotopic
chains of semi-magic even-even tin and lead nuclei within a version
of the self-consistent TFFS which  is very similar to that in
\cite{mu1,mu2}. The only difference  is  the use of a new version
DF3-a \cite{Tol-Sap} of the original DF3 functional \cite{Fay5,Fay}.
It differs from DF3 only by the spin-orbit parameters
$\varkappa,\varkappa'$ and the first  harmonics $g_1,g_1'$ of the
spin LM amplitude. The corresponding force components determine
mainly the spin-orbit doublet splitting and are of especial
importance for high $j$-levels. The last two terms modify  the
spin-orbit component of the mean field together with the spin-orbit
density $\rho_{sl}(r)$ which is determined with particles of
partially filled spin-orbit doublets. As it is well known, the first
spin harmonics contribute to the mean field in a combination with
tensor forces, see e.g. \cite{KhS}. Therefore, they should be
considered as the effective first harmonic or, on equal footing, the
effective tensor force. It should be mentioned that in both the
functionals DF3 and DF3-a the parameter $g_1$ is taken equal to
zero, thus, the spin-orbit splitting is determined with the set of
three parameters, $\varkappa,\varkappa',g_1'$. For brevity, we will
name them just spin-orbit. The functional DF3-a is characterized by
a rather strong effective tensor force.

In this paper we concentrate on quadrupole moments. We limit
ourselves to semi-magic nuclei, where the one-quasiparticle
approximation is expected to provide a good accuracy. In addition to
two isotopic chains considered in \cite{BE2}, we calculate
quadrupole moments of odd neighbors of even isotones with $N=50$ and
$N=82$ and of four odd neighbors of the magic nucleus $^{40}$Ca. We
compare predictions of two functionals, DF3 and DF3-a, which  have
different spin-orbit components. It is important for the problem
under consideration as quadrupole moments $Q_{\lambda}$ are
proportional to the Bogolyubov factor \cite{solov}: \beq
u^2_{\lambda}-v^2_{\lambda}=(\eps_{\lambda}-\mu)/E_{\lambda},\label{Bog}\eeq
where
$E_{\lambda}=\sqrt{(\eps_{\lambda}-\mu)^2+\Delta_{\lambda}^2}$, with
obvious notation. In the quadrupole moments problem we deal with the
ground state of an odd nucleus
 or with very low-lying excited state when often the inequality
$ |\eps_{\lambda}-\mu| \ll \Delta_{\lambda}$ is true. In such a
situation, the quadrupole moment value is very sensitive to the
single-particle energy $\eps_{\lambda}$. In its turn, the latter is
very sensitive to the spin-orbit parameters of the EDF. This could
help in fixing these parameters
 not known sufficiently well up to now.  The reason is that nuclear masses
and radii used mainly for finding effective force parameters are
not-sensitive to such parameters as $\kappa',g_1,g_1'$. Note that
the relevance of the spin-orbit and effective tensor force to other
low-energy nuclear phenomena was discussed recently in Refs.
\cite{Dug1,Dug2,Bald4}.
 In this paper, we consider only the surface kind
of pairing as motivated by previous research of Refs. \cite{Fay,Bald2,Bald4}.

\section{Brief calculation scheme}
The calculation scheme of the self-consistent TFFS based on the EDF method
by Fayans et al. is described in detail in Ref. \cite{BE2}. Here we summarize
several formulas which are necessary for understanding the main
ingredients of the approach. The EDF method by Fayans et al. \cite{Fay1,Fay5,Fay}
is a generalization for superfluid systems of the original  Kohn--Sham EDF method \cite{KSh}.
For condensed matter case, similar generalization was carried out in Ref. \cite{oliveira}.
In this method, the ground
state energy of a nucleus is considered as a functional of normal
and anomalous densities, \beq E_0=\int {\cal E}[\rho_n({\bf
r}),\rho_p({\bf r}),\nu_n({\bf r}),\nu_p({\bf r})] d^3r.\label{E0}
\eeq

Within the TFFS, the static quadrupole moment $Q_{\lambda}$ of an odd nucleus
with the  odd nucleon in the state $\lambda$  can be
found in terms of the diagonal matrix element $ \langle\lambda|
V(\omega =0)|\lambda\rangle$  of the effective field $V$ in
the static external field  $ V_0 = \sqrt{16\pi /5} r^2 Y_{20}$.
In systems with pairing correlations, equation  for the effective
field can be written in a compact form as \beq  \label{Vef_s}{\hat
V}(\omega)=e_q {\hat V}_0(\omega)+{\hat {\cal F}}  {\hat A}(\omega) {\hat V}(\omega),
\eeq where $e_q$ is the local quasiparticle charge with respect to the external field $V_0$ and
all other terms  are  matrices. In the
standard TFFS notation \cite{AB1}, we have:
\beq {\hat V}=\left(\begin{array}{c}V
\\d_1\\d_2\end{array}\right)\,,\quad{\hat
V}_0=\left(\begin{array}{c}V_0
\\0\\0\end{array}\right)\,,
\label{Vs} \eeq

\beq {\hat {\cal F}}=\left(\begin{array}{ccc}
{\cal F} &{\cal F}^{\omega \xi}&{\cal F}^{\omega \xi}\\
{\cal F}^{\xi \omega }&{\cal F}^\xi  &{\cal F}^{\xi \omega }\\
{\cal F}^{\xi \omega }&{\cal F}^{\xi \omega }& {\cal F}^\xi \end{array}\right), \label{Fs} \eeq

\beq {\hat A}(\omega)=\left(\begin{array}{ccc} {\cal L}(\omega) &{\cal M}_1(\omega)
&{\cal M}_2(\omega)\\
 {\cal O}(\omega)&-{\cal N}_1(\omega) &{\cal N}_2(\omega)\\{\cal O}(-\omega)&-{\cal N}_1(-\omega) &
 {\cal N}_2(-\omega)
\end{array}\right)\,,
\label{As} \eeq where ${\cal L},\; {\cal M}_1$, and so on stand
for integrals over $\eps$ of the products of different
combinations of the Green function $G(\eps)$ and two Gor'kov
functios $F^{(1)}(\eps)$ and $F^{(2)}(\eps)$. They can be found in
\cite{AB1}. In our case, the local charges in Eq. (\ref{Vef_s}) are $e_q^p=1,\; e_q^n=0$.

Isotopic indices in Eqs. (\ref{Vs}-\ref{As}) are omitted for
brevity. The explicit form of the above equations
 is written down for the case of the electric ($t$-even)
symmetry we deal with.
 In Eq. (\ref{Fs}), ${\cal F}$ is the usual LM amplitude,
\beq {\cal F}=\frac {\delta^2 {\cal E}}{\delta \rho^2}, \label{LM}\eeq
${\cal F}^{\xi}$ is the density-dependent effective pairing interaction,
\beq {\cal F}^{\xi}(\rho)=\frac {\delta^2 {\cal E}}{\delta \nu^2}, \label{EPI}\eeq
and the amplitudes ${\cal F}^{\omega \xi}={\cal F}^{\xi
\omega}$ stand for the mixed second derivatives, \beq {\cal
F}^{\omega \xi}=\frac {\delta^2 {\cal E}}{\delta \rho \delta \nu}.
\label{LMxi} \eeq  In the case of volume pairing, one has ${\cal
F}^{\omega \xi}=0$, whereas for the case of surface pairing we deal the amplitude
${\cal F}^{\omega \xi}$ is non-zero and should be taken into account
when Eqs. (\ref{Vs}-\ref{As}) are solved. As the analysis of Ref. \cite{BE2} shows,
the component $V$ of the vector ${\hat V}$, as a rule, dominates. However, the fields
$d_1,d_2$  also contribute, and sometimes significantly, to the value of $Q_{\lambda}$.

The final expression for the quadrupole moment of an odd nucleus
is as follows \cite{AB1,solov}

 \beq\label{Qlam}
 Q^{p,n}_{\lambda} =  (u^2_{\lambda}-v^2_{\lambda}) V^{p,n}_{\lambda},
 \eeq
 where $u_{\lambda}$, $v_{\lambda}$ are the Bogolyubov coefficients and
 \beq\label{Vlam}
 V_{\lambda} =  -\frac{2j-1}{2j+2} \int  V(r) R_{nlj}^2(r)
 r^{2}dr.
 \eeq
It should be noted that we use here, just as in Ref. \cite{BE2},
 the diagonal in $\lambda$ approximation for the gap, $\Delta_{\lambda,\lambda'}=
 \Delta_{\lambda} \delta_{\lambda,\lambda'}$ where the Bogolyubov set
 of equations is reduced to the BCS scheme. In the general case,
the relation for the quadrupole moment similar to Eq. (\ref{Qlam}) is
more complicated.
The $j$-dependent factor in (\ref{Vlam}) appears due to the angular
integral \cite{BM1}. For $j>1/2$ it is always negative. For odd
neighbors of a magic nucleus the ``Bogolyubov'' factor in
(\ref{Qlam}) reduces to 1 for a particle state and to $-1$ for a
hole one.
If the odd nucleon belongs to the superfluid component,  the
factor $(u^2_{\lambda}-v^2_{\lambda})$ in
Eq. (\ref{Qlam}) becomes non-trivial, see Eq. (\ref{Bog}). It changes permanently
depending on the state $\lambda$ and the nucleus under
consideration. Note that in the case of magnetic moments the factor
of $(u^2_{\lambda}+v^2_{\lambda})=1$ appears \cite{solov} in the relation
analogous to (\ref{Qlam}). In our case, this factor
determines the sign of the quadrupole moment.
It depends essentially
on values of the single-particle basis energies $\eps_{\lambda}$
reckoned from the chemical potential $\mu$ as.
Keeping in mind such sensitivity, we found this quantity for a given
odd nucleus $(Z,N+1)$ or $(Z+1,N)$,  $N,Z$ even, with taking into account the
blocking effect in the pairing problem \cite{solov} putting the odd
nucleon to the state $\lambda$ under consideration. For the
$V_{\lambda}$ value in Eq. (\ref{Qlam}) for superfluid nuclei, we used the half-sum of
these values in two neighboring even nuclei.

\begin{figure*}[ht!]
\centerline {\includegraphics [width=160mm]{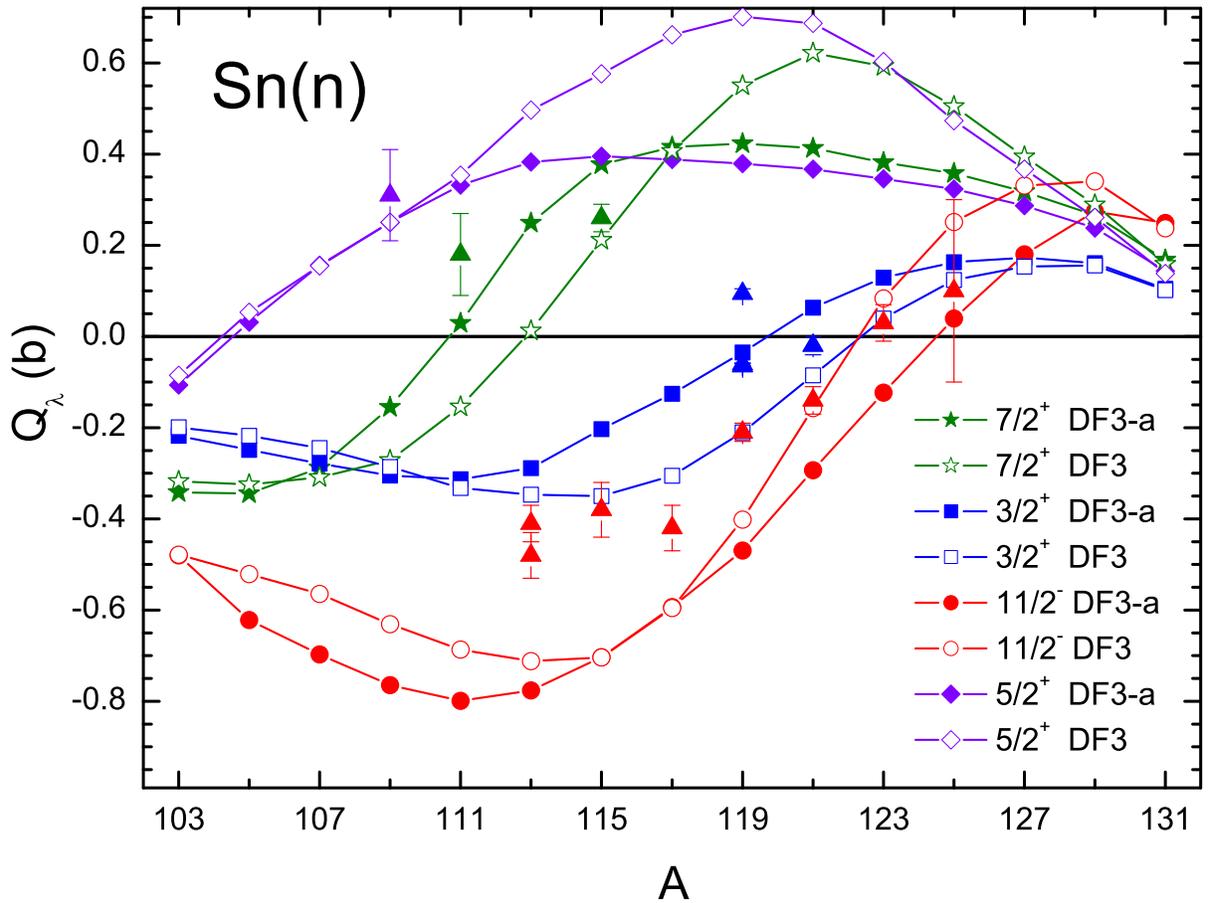}} \vspace{2mm}
\caption{ (Color online) Quadrupole moments of odd tin isotopes   with DF3 and DF3-a functionals.
 Triangles
with the bars indicate experimental data \cite{Stone}.}
\label{fig:QSn_n}
\end{figure*}

\begin{figure}[tbp]
\centerline {\includegraphics [width=80mm]{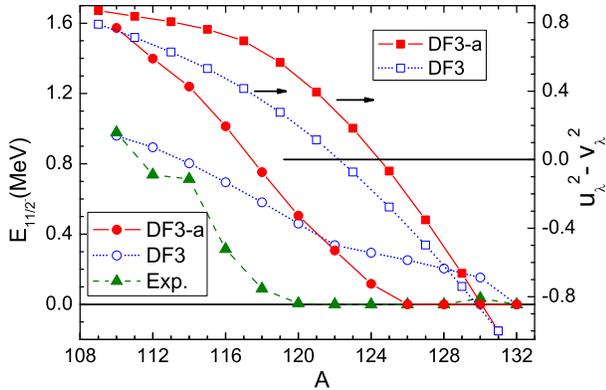}} \vspace{2mm}
\caption{  (Color online) For the $h_{11/2}$ level  in Sn isotopes:
the energy $E_{\lambda}$ accounted from the ground state(the
left-down legend; left axis) and values of the Bogolyubov
$(u_{\lambda}^2-v_{\lambda}^2)$ factor (right-up legend; right
axis).} \label{fig:lev_h11-2}
\end{figure}

\begin{figure}[tbp]
\centerline {\includegraphics [width=80mm]{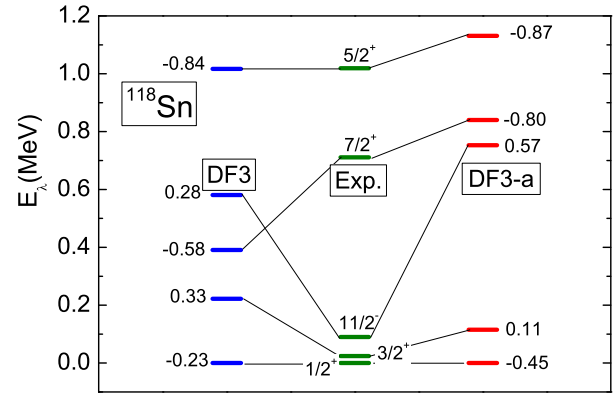}} \vspace{2mm}
\caption{  (Color online) Single-particle spectrum of the $^{118}$Sn nucleus. Value of the Bogolyubov $(u_{\lambda}^2-v_{\lambda}^2)$ factor is given
near each theoretical crossbar.}
\label{fig:lev_Sn118}
\end{figure}

\section{Calculation results}

The set of nuclei analyzed includes, in addition to odd neighbors
of two chains of even isotopes of tin ($Z=50$) and lead ($Z=82$) considered
in \cite{BE2}, odd neighbors of even isotones with $N=50$ and $N=82$. To check
the method for lighter nuclei, we calculated also quadrupole moments of odd
neighbors of the magic $^{40}$Ca nucleus.
 We concentrate on
 one of the two versions of pairing interaction considered
in \cite{BE2}, the surface one, as it was favored by that
investigation.
 On the other hand, we consider two versions
of the EDF. In addition to the DF3-a functional used in \cite{BE2} we made
alternative calculations with the original DF3 functional of \cite{Fay5,Fay}.
All parameters of the two functionals are given in \cite{BE2}.

Let us begin by analysing nuclei with odd neutron numbers.
Quadrupole moments of the odd Sn isotopes are displayed in Fig.\ref{fig:QSn_n}. Predictions
for the two versions of the functional, DF3 and DF3-a with different spin-orbit parameters,
 are compared with each other and with
the experimental data. In average, the results of  both  versions
reasonably agree with the data. Each theoretical
curve crosses the zero line due to vanishing of the corresponding Bogolyubov factor
in Eq. (\ref{Qlam}). The analogous experimental curve could be approximately displayed
only for the $11/2^-$-state. We see that it crosses the zero close to  both theoretical
curves, the difference between those is $\delta A\simeq 2$.

Before  analysing  the results for both versions of the EDF in
detail, it is instructive to discuss the $A$-dependence of the
position of the state $1h_{11/2}$ and compare it with the
experimental one. For the tin chain, this ``intruder'' state plays a
special role due to the high $j$ value and closeness to the Fermi
level. Indeed, the propagator ${\cal L}(\omega=0)$ of Eq. (\ref{As})
playing the main role in the TFFS equation for the effective field
$\hat{V}_{2^+}(\omega=0)$ contains  diagonal elements ${\cal
L}_{\lambda\lambda}=-\Delta_{\lambda}^2/(2E_{\lambda}^3)$. Their
contribution is proportional to the $(2j_{\lambda}+1)$ factor. If
the energy $E_{\lambda}(h_{11/2})$ is small, the contribution of
this level  to ${\cal L}$ can be rather big. Therefore a change in
the position of this level may change the result for the quadrupole
moment significantly.

As it was explained above, the DF3 and DF3-a functionals differ with spin-orbit parameters only, the difference
being important mainly for  high $j$-levels. The $1h_{11/2}$ level position accounted from the ground
state in Sn isotopes, $E_{\lambda}-\mu$, for both the functionals for even tin isotopes is displayed in
Fig. \ref{fig:lev_h11-2} in comparison with the experimental excitation spectrum in neighboring odd nuclei.
The corresponding values of the Bogolyubov factor $u_{\lambda}^2-v_{\lambda}^2$ are shown in the right-up
sector of the figure.
 Dealing with an even-even $^{A}$X nucleus, say, even
$^{A}$Sn isotope, there is a dilemma how to interpret   a state $|\lambda\rangle$ under consideration, either the hole state (i.e. the
excitation of the $^{A-1}$Sn nucleus) or the particle one (the
excitation of the $^{A+1}$Sn nucleus). We use a simple ansatz: the
state $|i \rangle$ is considered as a hole state if the inequality
$(u_{\lambda}^2-v_{\lambda}^2)<0$ takes place and as a particle state otherwise.  Note that
in the case of  $(u_{\lambda}^2-v_{\lambda}^2)\simeq 0$ the difference between the
particle and hole energies is, as a rule, quite small. Fortunately, for the main
 part of the chain there is
no contradiction between both the functionals in the ``particle/hole'' interpretation of
the  $1h_{11/2}$ state.
In the left part, up to  $^{120}$Sn nucleus, we deal with particle states and, correspondingly,
for each $A$ value here the green triangle shows the excitation energy of the $(A+1)$-th isotope.
On the right side, $A\ge 120$ experimental data show the ground state $11/2^-$, or almost
ground state in  $^{130,132}$Sn isotopes. Therefore,
a formal contradiction in this point for the  $^{122}$Sn doesn't work.
We see that both functionals predict $A$-dependence of the excitation energy of the
$h_{11/2}$ level qualitatively similar to the experimental one, falling in the value
with $A$ increasing, but both theoretical curves are higher than the experimental one.
Beginning from $A=124$, the DF3-a points practically coincide with the experimental ones.
The DF3-a curve is above the DF3 curve to the left of the crossing point between
$A=120$ and $A=122$, to the right, the position is inverse. For almost all the $A$ values
the inequality $(u_{\lambda}^2-v_{\lambda}^2)_{\rm DF3}<(u_{\lambda}^2-v_{\lambda}^2)_{\rm DF3-a}$
is valid. Absolute values
of the Bogolyubov factor for these functionals differ often significantly, as a rule,  more
than the quadrupole moment values, see Fig. \ref{fig:QSn_n} and Table \ref{tab:Q_n}.
Analysis shows that it occurs due to the opposite effect in the matrix elements $V_{\lambda}$
of the effective field.

\begin{figure}[tbp]
\centerline {\includegraphics [width=80mm]{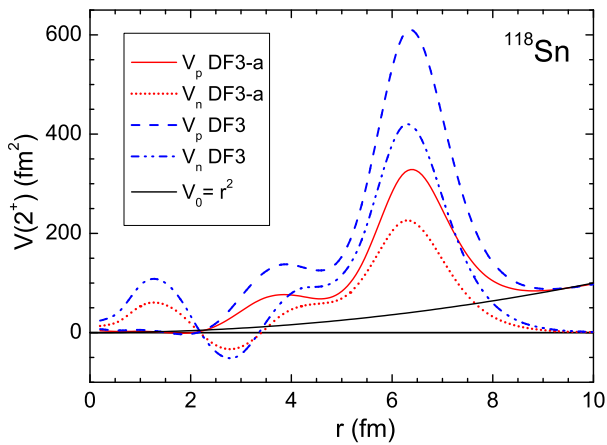}} \vspace{2mm}
\caption{ (Color online) Effective field $V(r)$ in $^{118}$Sn nucleus with DF3 and DF3-a functionals.}
\label{fig:V_Sn118}
\end{figure}

Let us analyze this point in more detail for the nucleus $^{118}$Sn which is in the
middle of the tin chain. The excitation spectrum for this nucleus is displayed in Fig. \ref{fig:lev_Sn118}.
According to the ansatz formulated above, position of experimental levels for $3/2^+,11/2^-$ states are
taken from the spectrum of the $^{119}$Sn nucleus, that of $7/2^+,5/2^+$ states, from $^{117}$Sn isotope.
Values of the Bogolyubov $u^2_{\lambda}-v^2_{\lambda}$ are also calculated in corresponding odd nuclei with
the blocking effect taken into account as described in Chapter 2.  In this nucleus positions of $11/2^-$  state
for both functionals are rather close to each other as this nucleus is in the vicinity of the crossing point
in Fig. \ref{fig:lev_h11-2}. However energies of the $7/2^+$ state, also with rather big momentum value, are
strongly different. On the whole, the density of states at the Fermi surface for the DF3 functional is higher
than for DF3-a one. This difference is, in fact, significant if one takes into account the factor of
$(2j+1)$ with which each $j$-state comes to this quantity.
To avoid misleading, we note that the comparison of the theoretical spectrum
in the even-even $A$-nucleus which is  the basis for the QRPA-like calculation
with the experimental excitation spectra of neighboring odd nuclei, as is made in Figs. \ref{fig:lev_h11-2},\ref{fig:lev_Sn118}, is,
in general, rather approximate operation. In heavy nuclei,  corrections to  the mean field single-particle
spectra appear due to particle-phonon coupling. Phenomenological functionals, as DF3 or DF3-a,
or any SHF functional, take into account such corrections only on average. For quantitative description
of single-particle spectra, all fluctuations,
from one nucleus to another  and from one state to another, should be calculated
explicitly.

\begin{table}[tbp]
\caption{Characteristics of single-particle states in $^{118}$Sn nucleus.}

\begin{tabular}{ c c c c c c c}
\hline \hline
$\lambda$ &$u^2_{\lambda}-v^2_{\lambda}$& $V_{\lambda}$& $Q_{\lambda}$&
$u^2_{\lambda}-v^2_{\lambda}$ &$V_{\lambda}$& $Q_{\lambda}$ \\
\hline
&&DF3&&&DF3-a& \\

2$d_{5/2}$ &-0.84&-0.838&+0.704  &-0.87&-0.449&+0.390 \\
1$g_{7/2}$ &-0.58&-0.954&+0.553  &-0.80&-0.549&+0.439 \\
1$h_{11/2}$&0.28 &-1.454&-0.407  &0.57 &-0.826&-0.471 \\
2$d_{3/2}$ &0.33 &-0.636&-0.210  &0.11 &-0.341&-0.038 \\
\hline \hline
\end{tabular}
\label{tab:lev-Sn118}
\end{table}

To complete the analysis of quadrupole moments of the odd neighbors of  $^{118}$Sn,
 we presented in Table \ref{tab:lev-Sn118}
 matrix elements  $V_{\lambda}$ entering Eq. (\ref{Qlam}) and quadrupole
moments $Q_{\lambda}=(u^2_{\lambda}-v^2_{\lambda})V_{\lambda}$. Note
that there is a small difference between these values of
$Q_{\lambda}$ and those in Fig. \ref{fig:QSn_n} and Table
\ref{tab:Q_n} as in our systematic calculations of quadrupole
moments we used in this product for an odd nucleus under
consideration the half-sum of  $V_{\lambda}$ values in neighboring
even nuclei. We see that absolute values of $V_{\lambda}$ for the
DF3 functional are almost two times bigger than for the DF3-a
functional used by us in the previous article \cite{BE2}. However,
in three cases of four the difference of Bogolyubov factors is
opposite and compensates significantly too big values of
$V_{\lambda}$  for the DF3 functional. This compensation is almost
complete for $7/2^+$ and $11/2^-$ states, the corresponding
quadrupole moments for two functionals almost coincide in vicinity
of $^{118}$Sn. This is not the case for the $5/2^+$ state where two
Bogolyubov factors almost equal to each other and $V_{\lambda}$[DF3]
almost twice bigger of $V_{\lambda}$[DF3-a)]. A strong enhancement
of the effective field $\hat{V}(r)$ for the DF3 functional
  in comparison with DF3-a one is shown in Fig. \ref{fig:V_Sn118}.
It is seen that the surface maxima of $V_p(r)$ and
 $V_n(r)$  functions are almost twice higher in the DF3 case. The higher density
of states at the Fermi surface
 discussed above is causing this effect.

\begin{figure}[tbp]
\centerline {\includegraphics [width=80mm]{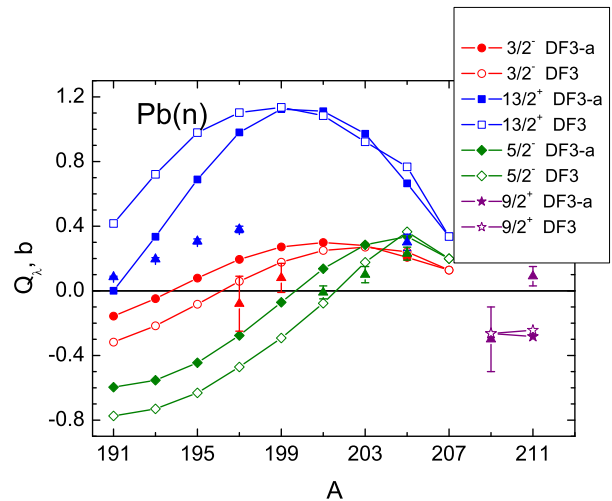}} \vspace{2mm}
\caption{  (Color online) Quadrupole moments of odd lead isotopes with DF3 and DF3-a functionals. Triangles
with the bars indicate experimental data \cite{Stone}.}
\label{fig:Qpb_n}
\end{figure}

\begin{figure}[tbp]
\centerline {\includegraphics [width=80mm]{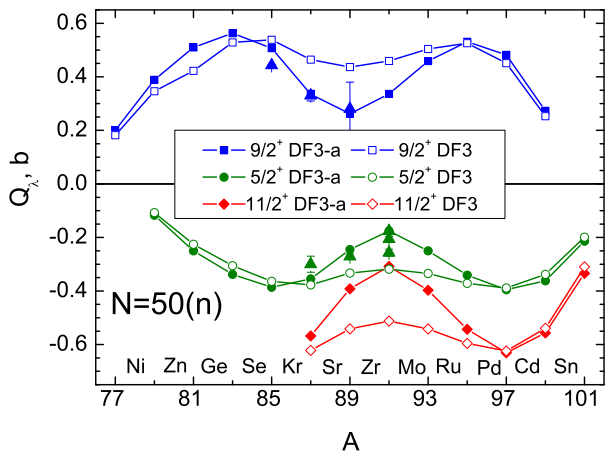}} \vspace{2mm}
\caption{  (Color online) Quadrupole moments of odd-neutron neighbors of even $N=50$ isotones
  with DF3 and DF3-a functionals.  Triangles with bars indicate experimental data \cite{Stone}.}
\label{fig:N50_n}
\end{figure}

 For a deeper understanding of this enhancement effect, we calculated for the nucleus
  $^{118}$Sn  characteristics of the $2^+_1$ state with the DF3 functional,
  $\omega_2=0.913\;$MeV and $B(E2)=0.272\;$e b, and compare them with those calculated previously \cite{BE2} for
  the DF3-a functional,  $\omega_2=1.217\;$MeV and $B(E2)=0.172\;$e b. Note that the DF3-a predictions are in
  reasonable agreement with experimental data, thus the DF3 functional leads to too collective  $2^+_1$ state.
  It explains qualitatively an additional enhancement of the static quadrupole field for the DF3 functional.
  Indeed, at small external quadrupole field frequency $\omega$, the pole in the $\omega=\omega_2$ point term
  dominates in the effective field,
  \beq \hat{V}(\omega)=\frac {2\omega_s \hat{\chi}_s} {\omega^2-\omega_s^2} +
\hat{V}_R(\omega), \label{V-pole} \eeq
  where the regular term $\hat{V}_R(\omega)\simeq const$ at small $\omega$. The residue of the pole term
  is $\hat{\chi}_s=(\hat{V}_0\hat{A}\hat{g}^+_s)\hat{g_s}$, $\hat{g_s}$ being the
amplitude of creating the $s$-th ($2^+_1$ in
  our case) phonon \cite{AB1}. It can be easily shown that the relation $\chi_2\propto B(E2)$ is valid.
  If to neglect for an estimation the regular term in Eq. (\ref{V-pole}), we arrive
  to $V(\omega=0)\propto B(E2)/\omega_2$.
  After substituting to this relation the values given above, we find this ratio equal
  to 0.298 for the DF3 functional and
  0.141 for DF3-a. This explains the effect under discussion.

To evaluate the agreement with experiment quantitatively, we
calculate the mean theory-experiment difference
 \beq
\sqrt{\overline{(\delta Q)^2_{\rm rms}}}  =  \sqrt{\frac 1 {\cal N} \sum_{i} \left(Q^{\rm th}_i-
Q^{\rm exp}_i\right)^2},\label{rms}\eeq
with obvious notation. For the tin chain under consideration, ${\cal N}=11$,  we have
$ \sqrt{\overline{(\delta Q)^2_{\rm rms}}} = 0.167\;$e b for the DF3 functional
and $\sqrt{\overline{(\delta Q)^2_{\rm rms}}} = 0.164\;$e b
for the DF3-a one. Thus, for the DF3-a functional agreement is somewhat better.
A remark should be made concerning practical application of Eq. (\ref{rms}). All terms of the sum
come with the same weight independent on the experimental error of each measurement.
The reason is that, as calculations \cite{BE2} have shown, our theoretical accuracy for
quadrupole moments is not better than 0.1 e b. Practically all  the
data we use for comparison have better accuracy therefore for our analysis they  are equivalent.
The nucleus $^{209}$Pb is the only exception where the experimental error is 0.2 e b. To take it into account,
we multiply this term with the factor $(0.1/0.2)^2=1/4$. If the table \cite{Stone} contains several different
data on the quadrupole moment under consideration, say 3, we take into account all of them in the
sum of (\ref{rms}) with the weight 1/3.

Let us go now to odd lead nuclei, Fig. \ref{fig:Qpb_n}. We see that
again there is a reasonable agreement with experiment for both the
functionals with the exception of   the intruder state $1i_{13/2}$. For the
latter, in the case of the DF3-a functional, a reasonable agreement with the data is found only
for two lightest isotopes $^{191,193}$Pb.  For other four nuclei with known values of $Q(1i_{13/2})$
disagreement is significant, the maximal value of $\delta Q=Q^{\rm th}-
Q^{\rm exp}$ for  $^{197}$Pb is 0.60 e b. For the DF3 functional the situation
is noticeably worse. Now the rms deviation of the theory from experiment, Eq. (\ref{rms}),
is  $ \sqrt{\overline{(\delta Q)^2_{\rm rms}}}[\rm DF3-a] = 0.277\;$e b and
 $ \sqrt{\overline{(\delta Q)^2_{\rm rms}}}[\rm DF3] = 0.386\;$e b. Now the quality
 of agreement became significantly worse than for tin chain for both
 functionals but the deterioration  is much stronger for the DF3 functional.

\begin{figure}[hb!]
\centerline {\includegraphics [width=80mm]{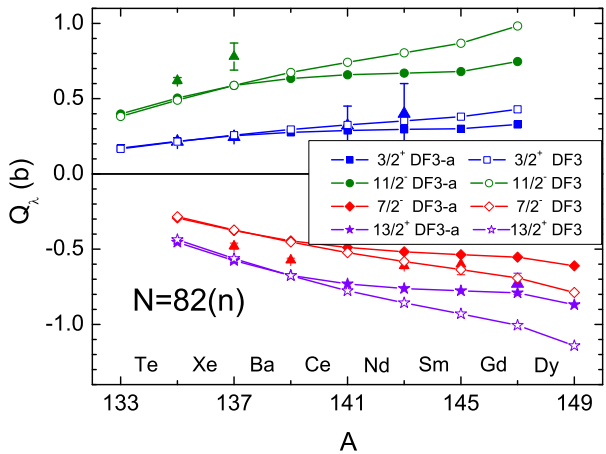}} \vspace{2mm}
\caption{ (Color online) Quadrupole moments of odd-neutron neighbors of even $N=82$
isotones  with DF3 and DF3-a functionals.  Triangles with the bars indicate experimental data \cite{Stone}.}
\label{fig:N82_n}
\end{figure}

 In Fig. \ref{fig:N50_n} quadrupole moments are displayed of odd-neutron
 neighbors, $N=50\pm1$, of even isotones with $N=50$. In this case, the
 proton-subsystem is superfluid and the neutron Bogolyubov factor in Eq. (\ref{Qlam})
 is $\pm1$. Agreement with the data is almost perfect for the DF3-a functional,
$ \sqrt{\overline{(\delta Q)^2_{\rm rms}}}[\rm DF3-a] = 0.041\;$e b, and only a little worse for
the DF3 functional, $ \sqrt{\overline{(\delta Q)^2_{\rm rms}}}[\rm DF3] = 0.116\;$e b.
For this chain we showed also quadrupole moments of the intruder state $1h_{11/2}$ although
none of them was measured. Contrary to tin and lead isotopic chains, none of the functions
$Q_{\lambda}(A)$ changes the sign.
The situation is similar for odd-neutron neighbors
of even isotones with $N=82$  presented in Fig. \ref{fig:N82_n}.
Again agreement with the data is very good for the DF3-a functional,
$ \sqrt{\overline{(\delta Q)^2_{\rm rms}}}[\rm DF3-a] = 0.093\;$e b, and again only a little worse for
the DF3 functional, $ \sqrt{\overline{(\delta Q)^2_{\rm rms}}}[\rm DF3] = 0.119\;$e b. In this case,
there is one experimental data for the intruder state $1i_{13/2}$ for which the DF3-value
practically coincides with experiment.

 Table  \ref{tab:Q_n} contains odd-neutron nuclei with known experimental quadrupole moments.
 We have seen that for each chain considered the DF3-a functional describes data better than
 the DF3 one. Therefore we present here only predictions of the DF3-a functional. The table
is divided in two parts. In the bottom,  the intruder states
$1h_{11/2}$ and $1i_{13/2}$ are collected. The rms deviation of the
theoretical predictions from the data calculated with Eq. \ref{rms}
for all 42 moments in this Table is rather big, $
\sqrt{\overline{(\delta Q)^2_{\rm rms}}}[\rm DF3-a] = 0.189\;$e b.
For the DF3 functional, the error is bigger, $
\sqrt{\overline{(\delta Q)^2_{\rm rms}}}[\rm DF3] = 0.240\;$e b.
Examining the table, we find that the main part of big deviations
$\delta Q$ is concentrated in the ``intruder'' part of the Table
containing 15 moments. If to make the calculation (\ref{rms}) only
for the latter, we find  $ \sqrt{\overline{(\delta Q)^2_{\rm
rms}}}[\rm intruder] = 0.269\;$e b. At last, rather moderate value
of the error will occur if we limit ourselves with quadrupole
moments of the rest of 27 ``normal'' states, $
\sqrt{\overline{(\delta Q)^2_{\rm rms}}}[\rm normal] =  0.125\;$e b.

In conclusion of this discussion, note that the formalism of developed pairing we use
with particle number conservation  on average, as is known \cite{part-numb},
works worse in the vicinity of magic numbers. For completeness, we include into
analysis nuclei $^{205}$Pb and $^{211}$Pb
``dangerous'' from this point of view. However, in this case it is natural
to expect significant errors induced
by this approximation.

\begin{table}[ht!]
\caption{Quadrupole moments $Q\;$(e\;b) of odd-neutron nuclei
in the state $\lambda$.
Theoretical values $Q_{\rm th}$ and differences $\delta Q=Q_{\rm th}-Q_{\rm exp}$
are given for the functional DF3-a.}

\begin{tabular}{lcccc }
\hline \hline nucleus  &$\lambda$  & $Q_{\rm exp}$
&\hspace*{1.5ex}$Q_{\rm th}$\hspace*{1.5ex}& $\delta Q$  \\

\hline
$^{39}$Ca &$1d_{3/2}$  &0.036(7)  &+0.040 &0.004 \\
          &            & 0.040(6) &       & 0.000 \\
$^{41}$Ca &$1f_{7/2}$ &-0.090(2)  &-0.078 &0.012\\
&                     &  -0.066(2)&       &-0.012  \\
&                     &-0.080(8)  &       &0.002\\
$^{85}$Kr  &$1g_{9/2}$  &  +0.443(3) &+0.507 &0.064  \\

$^{87}$Kr  &$2d_{5/2}$  &  -0.30(3)  &-0.355 &-0.06 \\

$^{87}$Sr  &$1g_{9/2}$  &  +0.33(2)  &+0.335& 0.01  \\

$^{89}$Sr  &$2d_{5/2}$  &  -0.271(9)  &-0.245  &-0.026 \\

$^{89}$Zr  &$1g_{9/2}$  &   +0.28(10) &+0.262  &-0.02 \\

$^{91}$Zr   &$2d_{5/2}$  &   -0.176(3) &-0.195 &-0.019 \\
            &            &   (-)0.257(13)       &     & 0.062 \\
            &            &   -0.206(10)         &    & 0.011 \\

$^{109}$Sn &$2d_{5/2}$ &+0.31(10)                  & +0.250 &-0.06 \\

$^{111}$Sn &$1g_{7/2}$ &+0.18(9)                   &+0.029   &-0.13\\

$^{115}$Sn &$1g_{7/2}^{\,*}$ &0.26(3)              &+0.377  &0.12\\

$^{119}$Sn &$2d_{3/2}^{\,*}$ &0.094(11) & -0.035   &-0.129  \\

           &                 &-0.065(5) &           & 0.030   \\

           &                 &-0.061(3) &           & 0.026   \\

$^{121}$Sn &$2d_{3/2}$ &-0.02(2)                 &+0.063 & 0.08  \\

$^{135}$Xe &$2d_{3/2}$ &    +0.214(7)               &+0.217& 0.003\\

$^{137}$Xe&  $2f_{7/2}$  &     -0.48(2)           &-0.376&0.10 \\

$^{137}$Ba&$2d_{3/2}$ &       +0.245(4)           &+0.254&0.009 \\

$^{139}$Ba& $2f_{7/2}$  &   -0.573(13)            & -0.445&0.128 \\

$^{141}$Nd &$2d_{3/2}$ &    +0.32(13)             &+0.289 &-0.03 \\

$^{143}$Nd&  $2f_{7/2}$  &  -0.61(2)              &-0.518 &0.09 \\

$^{143}$Sm &$2d_{3/2}$ &     +0.4(2)              &+ 0.296 &0.1\\

$^{145}$Sm&  $2f_{7/2}$  &  -0.60(7)              &-0.537 &0.06 \\

$^{197}$Pb &$3p_{3/2}$ &-0.08(17)                &+0.195 &0.27  \\

$^{199}$Pb &$3p_{3/2}$ &+0.08(9)                 &+0.272   & 0.19\\

$^{201}$Pb &$2f_{5/2}$ &-0.01(4)                 &+0.137  & 0.15\\

$^{203}$Pb &$2f_{5/2}$ &+0.10(5)                  &+0.284  & 0.18\\

$^{205}$Pb &$2f_{5/2}$ &+0.23(4)                 &+0.336   &0.09  \\

$^{209}$Pb &$2g_{9/2}$ &-0.3(2)                  &-0.264   &0.1\\

$^{211}$Pb &$2g_{9/2}$ &+0.09(6)                 &-0.283   &-0.37\\

\hline\hline

$^{113}$Sn &$1h_{11/2}^{\,*}$ &0.41(4)          &-0.776 & -0.37 \\
           &                  & 0.48(5)       &          & -0.30  \\

$^{115}$Sn &$1h_{11/2}^{\,*}$ &0.38(6)           &-0.703   &-0.32 \\

$^{117}$Sn &$1h_{11/2}^{\,*}$ &-0.42(5)          &-0.593& -0.17  \\

$^{119}$Sn &$1h_{11/2}^{\,*}$ &0.21(2)           &-0.469&  -0.25 \\

$^{121}$Sn &$1h_{11/2}^{\,*}$ &-0.14(3)          &-0.293  &-0.15 \\

$^{123}$Sn &$1h_{11/2}$ &+0.03(4)                &-0.123   &-0.15 \\

$^{125}$Sn &$1h_{11/2}$ &+0.1(2)                 &+0.039   & -0.1  \\

$^{135}$Xe &$1h_{11/2}^{\,*}$ &   +0.62(2)         &+0.504&0.12 \\

$^{137}$Ba &$1h_{11/2}^{\,*}$ &  +0.78(9)            &+0.588&-0.19 \\

$^{147}$Gd& $1i_{13/2}^{\,*}$ &  -0.73(7)          & -0.791 &-0.06 \\

$^{191}$Pb &$1i_{13/2}^{\,*}$ &+0.085(5)          &+0.0004 &-0.085   \\

$^{193}$Pb &$1i_{13/2}^{\,*}$ &+0.195(10)         &+0.335  &0.140  \\

$^{195}$Pb &$1i_{13/2}^{\,*}$ &+0.306(15)         &+0.689  &0.383  \\

$^{197}$Pb  &$1i_{13/2}^{\,*}$ &+0.38(2)          &+0.980  &0.60   \\

 $^{205}$Pb &$1i_{13/2}^{\,*}$& 0.30(5)           &+0.665  &0.37  \\

\hline \hline
\end{tabular}
\label{tab:Q_n}
\end{table}

\begin{figure}[ht!]
\centerline {\includegraphics [width=80mm]{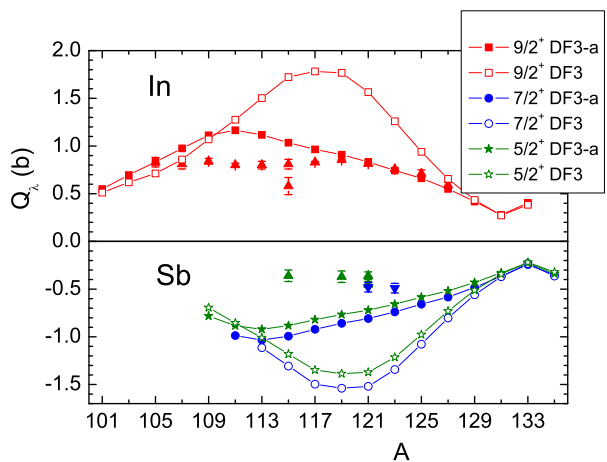}} \vspace{2mm}
\caption{ (Color online) Quadrupole moments of odd-proton neighbors of even tin
isotopes  with DF3 and DF3-a functionals.  Triangles
with the bars indicate experimental data \cite{Stone}.}
\label{fig:QSn_p}
\end{figure}

Let us go to odd-proton nuclei. Quadrupole moments of odd-proton neighbors
of even tin isotopes, those of In ($Z=49$) and Sb (Z=51) are displayed in
Fig. \ref{fig:QSn_p}. The Bogolyubov factor in (\ref{Qlam}) is now reduced
to -1 for the In chain and +1 for the Sb chain, thus, all difference between predictions
of the two functionals we use may come only from different values of the
matrix elements $V_{\lambda}$ of the effective field. We see that in the
middle part of both chains $|Q_{\lambda}|$ for the DF3 functional exceeds
that for the DF3-a in two times approximately. Remind that the effective
field is that in the even Sn core, the same as was used for odd tin isotopes.
Now the matrix elements are taken from the proton component  $V_p(r)$ whereas
previously we dealt with the neutron component $V_n(r)$. Both were displayed
in Fig. \ref{fig:V_Sn118} and we have seen that both of  them are approximately
two times bigger for the DF3 functional. This was caused by a too strong collective
$2^+_1$ state for this functional. For tin isotopes, this deficiency was partially
compensated with different values of the Bogolyubov factor, see Tab. \ref{tab:lev-Sn118}.
Now we see the effect of the effective field directly, without any distortions.
In the case of the DF3-a functional, the rms deviation is worse than for tin isotopes
but moderately,  $ \sqrt{\overline{(\delta Q)^2_{\rm rms}}}[\rm DF3-a] = 0.249\;$e b.
As to the DF3 functional, the error grows drastically,
$ \sqrt{\overline{(\delta Q)^2_{\rm rms}}}[\rm DF3] = 0.688\;$e b. There is a feeling
 that for the DF3-a functional the effective field $V_p(r)$ is also too strong and
the agreement will become better if it will be reduced by 20-30\%. Within the mean field
approach we use it can be achieved only with such variation of the central part of the EDF,
common to DF3 and DF3-a, which changes the LM amplitude ${\cal F}_{np}$.

\begin{figure}[ht!]
\centerline {\includegraphics [width=80mm]{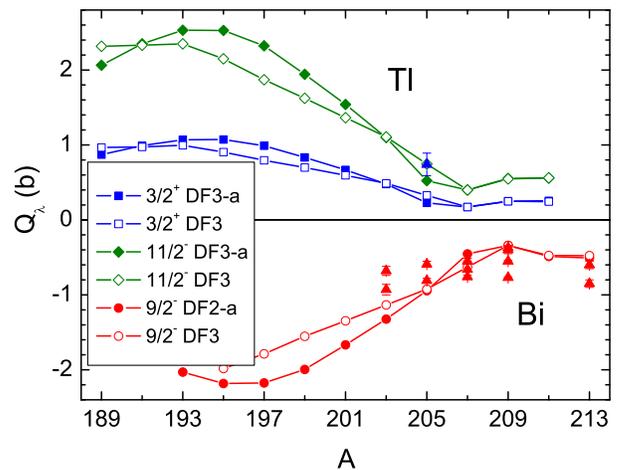}} \vspace{2mm}
\caption{ (Color online) Quadrupole moments of odd-proton neighbors of even lead
isotopes  with DF3 and DF3-a functionals.  Triangles
with the bars indicate experimental data \cite{Stone}.}
\label{fig:Qpb_p}
\end{figure}

Quadrupole moments of the odd neighbors of even lead isotopes are presented
in Fig. \ref{fig:Qpb_p}. We see that here the difference between DF3 and DF3-a
functionals is not big, especially in the right half of the figure where
some experimental data exist. Unfortunately, their number is limited,
only one for Tl and five for Bi. In the last case, there are two sets of the data \cite{Stone}.
Both the calculations agree with one of them and disagree with the other. In such a situation,
it is not worth to estimate the average disagreement numerically as it was made above.

\begin{figure}[tbp]
\centerline {\includegraphics [width=80mm]{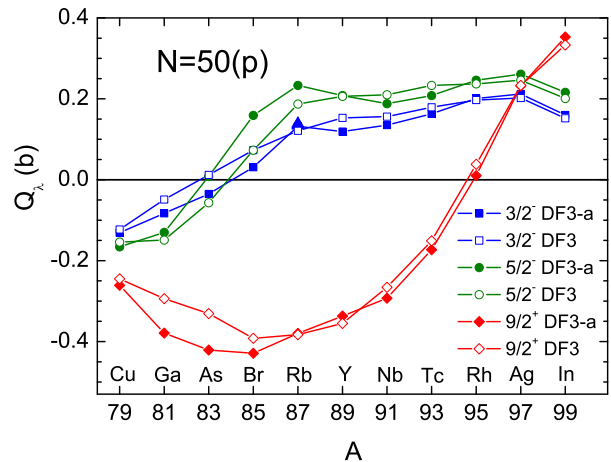}} \vspace{2mm}
\caption{ (Color online) Quadrupole moments of odd-proton $N=50$ isotones  with DF3 and DF3-a functionals.
 Triangles with the bars indicate experimental data \cite{Stone}.}
\label{fig:N50_p}
\end{figure}

Let us go to the chain of odd-proton with magic neutron number $N=50$, eleven
nuclei from $^{79}$Cu till $^{99}$In, see Fig. \ref{fig:N50_p}. Note that all of them,
except $^{89}$Y, are $\beta$-unstable,
many being far beyond the $\beta$-stability valley. Here we deal with proton pairing, and the
Bogolyubov factor in Eq. (\ref{Qlam}) varies strongly from one nucleus to another and depends on
the state $\lambda$. However, for this chain we obtain rather close predictions of both
functionals used. In this case, the difference between both functionals is not as strong as in
the case of the $1h_{11/2}$ intruder state in the tin isotopes.
Unfortunately, the experimental quadrupole moment is known for  one nucleus only, $^{87}$Rb.
For this, both theoretical predictions  practically coincide with the experimental values.

\begin{figure}[tbp]
\centerline {\includegraphics [width=80mm]{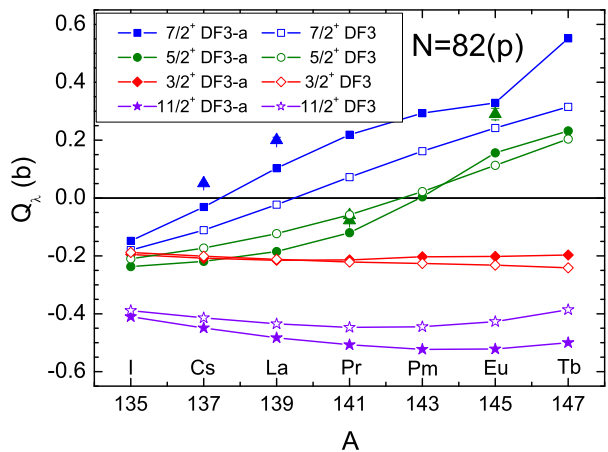}} \vspace{2mm}
\caption{ (Color online) Quadrupole moments of odd-proton $N=82$ isotones with DF3 and DF3-a functionals.
 Triangles with the bars indicate experimental data \cite{Stone}.}
\label{fig:N82_p}
\end{figure}

Quadrupole moments of the chain of odd-proton with magic neutron number $N=82$ are displayed
in Fig. \ref{fig:N82_p}. Here we deal with seven nuclei, from $^{135}$I till $^{147}$Tb. Among them
there are two stable ones, $^{139}$La and $^{141}$Pr, the others are $\beta$-unstable, but
only $^{135}$I lives
hours, the rest, days or years. For this chain, four quadrupole moments are known. For the DF3-a
functional, disagreement with the data doesn't exceed 0.1 e b. For the DF3 one, disagreement is
greater but not significantly. Comparison between predictions of the two functionals
shows very close results for the $2d_{3/2}$ and $2d_{5/2}$ states. It demonstrates close results
for the effective field $V_p(r)$. A stronger difference for the $1g_{7/2}$ state is due to the difference of
the Bogolyubov factors in Eq. (\ref{Qlam}) similar to that we have seen for the neutron $1g_{7/2}$ state,
see Fig. \ref{fig:lev_Sn118}. It is worth to mention that the difference between two predictions
for the proton intruder $1i_{13/2}$ state is significantly less than for the neutron one in the lead
isotopes, Fig. \ref{fig:Qpb_n}. More exactly, for neutrons the difference was strong for lead isotopes
 lighter of $^{200}$Pb and rather small, for heavier ones. Such behavior is explained with dynamics of
 variation of the spin-orbit density with filling the neutron shell which strongly influences the
 position of high $j$-levels.

All quadrupole moments of odd-proton nuclei with known experimental values are collected in Table \ref{tab:Q_p}.
Just as for neutrons, we concentrated on the DF3-a functional which
turned out  to be more successful than the DF3 one. For the DF3 functional, the average error for
protons is too big, $ \sqrt{\overline{(\delta Q)^2_{\rm rms}}}[\rm DF3] = 0.589\;$e b, whereas
$ \sqrt{\overline{(\delta Q)^2_{\rm rms}}}[\rm DF3-a]  = 0.254\;$e b. However, for the DF3-a functional
in average protons are also described worse than neutrons.
The main contribution to this deviation comes from In and Sb isotopes, odd neighbors
of even tin nuclei. As it was written above, it is the result of too strong quadrupole
field $V_{n,p}(r)$ even for DF3-a functional, not only for the DF3 one. For neutrons this drawback is
partially
hidden with multiplying by the Bogolyubov factor, see Table \ref{tab:lev-Sn118}, but
for protons it appears to the full extent.
In many cases, we feel, the main reason of disagreement is the neglect of
 corrections from the particle-phonon coupling. Although in semi-magic
nuclei this coupling is relatively small and can be accounted for with
perturbation theory, sometimes these corrections could be noticeable.
E.g. for the excited $3/2^+$ state in $^{205}$Tl a strong admixture of
the $(2^+\bigotimes 1/2^+)_{3/2}$ state may be expected. It could explain the very strong
deviation $\delta Q$ in this case.

\begin{table}[b]
\caption{Quadrupole moments $Q\;$(b) of odd-proton nuclei
in the state $\lambda$.
Theoretical values $Q_{\rm th}$ and differences $\delta Q=Q_{\rm th}-Q_{\rm exp}$
are given for the functional DF3-a.}

\begin{tabular}{l c c c  c}
\hline \hline nucl.  &$\lambda$  & $Q_{\rm exp}$ &\hspace*{1.ex} $Q_{\rm
th}^{\rm DF3-a}$\hspace*{1.ex}&$\delta Q$(DF3-a)\\

\hline
$^{39}$K & $1d_{3/2}$&0.0585(6)     &0.069      &0.010  \\

$^{41}$Sc & $1f_{7/2}$&-0.156(3)    &-0.139     & 0.017 \\
&                     &0.120(6)    &            &-0.019\\
&                     &0.168(8)  &              & 0.029\\

$^{87}$Rb & $2p_{3/2}$ & +0.134(1)    &+0.132  &-0.002\\
&                      &  +0.138(1)   &        &-0.006\\

$^{105}$In & $1g_{9/2}$& +0.83(5) &   +0.833    & 0.00 \\

$^{107}$In & $1g_{9/2}$& +0.81(5) &+0.976       &0.17   \\

$^{109}$In & $1g_{9/2}$& +0.84(3) &+1.113       & 0.27 \\

$^{111}$In & $1g_{9/2}$& +0.80(2) &+1.165       &0.36 \\

$^{113}$In & $1g_{9/2}$& +0.80(4) &+1.117       &0.32\\

$^{115}$In & $1g_{9/2}$& +0.81(5) & +1.034      &0.22  \\

&                       & 0.58(9)&              &0.45\\

$^{117}$In & $1g_{9/2}$& +0.829(10)& +0.965     & 0.136  \\

$^{119}$In & $1g_{9/2}$& +0.854(7) &+0.909      & 0.055\\

$^{121}$In & $1g_{9/2}$& +0.814(11) &+0.833     &0.019 \\

$^{123}$In & $1g_{9/2}$& +0.757(9)  &+0.743     &-0.014   \\

$^{125}$In & $1g_{9/2}$& +0.71(4)   &+0.663     &-0.05 \\

$^{127}$In & $1g_{9/2}$& +0.59(3)   &+0.550    &-0.04  \\

$^{115}$Sb & $2d_{5/2}$& -0.36(6)   &-0.882   & -0.52       \\

$^{119}$Sb & $2d_{5/2}$& -0.37(6)   &-0.766   &-0.40 \\

$^{121}$Sb & $2d_{5/2}$& -0.36(4)   & -0.721 & -0.36 \\
&                      & -0.45(3)   &             & -0.27\\

$^{123}$Sb  & $1g_{7/2}$& -0.49(5)  &-0.739 & -0.25    \\

$^{137}$Cs  & $1g_{7/2}$&+0.051(1)    &-0.031  &-0.080 \\

$^{139}$La &  $1g_{7/2}$& +0.20(1)    &+0.103  &-0.10 \\

$^{141}$Pr &   $2d_{5/2}$& -0.077(6)  &-0.120  &-0.043 \\
&                        &  -0.059(4)  &       &-0.061\\

$^{145}$Eu &   $2d_{5/2}$&   +0.29(2)  &+0.156 &-0.13 \\

$^{205}$Tl &   $3d_{3/2}^*$ &+0.74(15)   & +0.227 &-0.51 \\

$^{203}$Bi  & $1h_{9/2}$ & -0.93(7) & -1.323 &-0.39 \\
            &           &  -0.68(6) &        & -0.64 \\
$^{205}$Bi  & $1h_{9/2}$& -0.81(3)  & -0.945  &-0.14 \\
            &           & -0.59(4)  &         &-0.36 \\
$^{207}$Bi  & $1h_{9/2}$& -0.76(2)  & -0.454  &0.31 \\
            &           & -0.55(4)  &        &0.10 \\
            &           &-0.60(11)  &        &0.15 \\

$^{209}$Bi  & $1h_{9/2}$& -0.516(15)   & -0.342 & 0.18  \\
&                       & -0.37(3)     &        & 0.03\\
&                       &  -0.55(1) &           & 0.21\\
&                       &  -0.77(1) &           &  0.43 \\
&                       &  -0.40(5) &           &  0.06 \\
&                       &  -0.39(3) &           &  0.05 \\

$^{213}$Bi  & $1h_{9/2}$& -0.83(5)  &-0.508   & 0.32 \\
            &           & -0.60(5)  &         & 0.09 \\

\hline \hline
\end{tabular}
\label{tab:Q_p}
\end{table}

\section{Conclusions}
 Quadrupole moments  of odd neighbors of semi-magic lead and tin
isotopes and $N=50,N=82$ isotones are calculated within the
self-consistent TFFS on the base of the Energy Density Functional by
Fayans et al. Two sets of the EDF  parameters are used fixed
previously. Namely, the functional DF3 \cite{Fay,Fay5} is used and
its  modification   DF3-a  \cite{Tol-Sap}. The latter is obtained
with a small variation  in the initial functional DF3 of the
spin-orbit terms and rather strong, of the effective tensor term.
The latter influences the spin-orbit splitting in nuclei with
partially filled spin-orbit doublets where the spin-orbit density
$\rho_{sl}$ is non-zero. Due to different spin-orbit splitting, the
position of high $j$-levels and all the single-particle spectrum in
the vicinity of Fermi level are often different for two functionals
under consideration. The complete set of the QRPA-like TFFS
equations for the effective field is solved in a self-consistent
way.

Recently, we used this method \cite{BE2} to calculate
characteristics of the $2^+_1$ states in even tin and lead isotopes
with the use of the DF3-a functional. In this work, quadrupole
moments were calculated as well of odd tin and lead isotopes and
odd-proton neighbors, In and Sb isotopes in the first case and Tl
and Bi isotopes in the second one. Strong sensitivity of quadrupole
moments of odd-neutron tin and lead isotopes to the single-particle
spectrum in the vicinity of the  Fermi surface was found in
\cite{BE2} due to the Bogolyubov factor
$u_{\lambda}^2-v_{\lambda}^2=(\eps_{\lambda}-\mu)/E_{\lambda}$.

To examine this effect in more detail, in the present paper we
calculated  quadrupole moments with the DF3 functional. These calculations
confirmed the observation of Ref. \cite{BE2}. In addition, it was found that the
static quadrupole effective field $V(r)$, the main ingredient of the formula for
the quadrupole moment $Q_{\lambda}$, also can be significantly different for these
two functionals, again due to different level structure in the vicinity of the Fermi level.
This occurs in the middle part of the tin isotopes where the DF3 effective field
approximately two times stronger than the DF3-a one. The latter looks more realistic
as follows from the analysis of the characteristics $\omega_2$ and BE2 in even tin
isotopes.{\footnote{In fact, such analysis was made for $^{118}$Sn nucleus.}} For odd
tin isotopes, a too strong effective field is hidden partially with smaller values
of the Bogolyubov factor, but the situation is worse for odd-proton neighbors, In and Sb
isotopes, where the DF3 functional leads to a strong disagreement with experiment.
For other chains considered the difference between the predictions of both functionals is less, but
on the whole, the DF3-a functional describes quadrupole moments  better than the DF3 one.

For the DF3-a functional, the agreement, on average, can be
considered as reasonable. For 42 quadrupole moments of odd-neutron
nuclei, the average disagreement between theory and experiment is
not so small, $ \sqrt{\overline{(\delta Q)^2_{\rm rms}}} = 0.189\;$e
b. However, it is concentrated mainly in 15 intruder states for
which we have  $ \sqrt{\overline{(\delta Q)^2_{\rm rms}}}[\rm
intruder] = 0.269\;$e b. For the rest of
 27 ``normal'' states, the disagreement is rather moderate
 $ \sqrt{\overline{(\delta Q)^2_{\rm rms}}}[\rm normal] =  0.125\;$e b.

For protons, agreement is worse. Leaving aside specific reasons in several cases,
we see the main source of disagreement here, just as for neutrons, in
neglecting the effects  of the coupling of single-particle degrees
of freedom with phonons, see e.g. \cite{mu1,mu2,kaev2011}.

The comparison between two functionals favors DF3-a  which has a rather strong effective tensor force.
For further improvements,  phonon contributions should be taken into account
which would pave the way to extend the present investigation
to nuclei with both non-magic proton and neutron subsystems.

\section{Acknowledgment}
The work was partly supported by the DFG
and RFBR Grants Nos.436RUS113/994/0-1 and 09-02-91352NNIO-a, by the
Grants NSh-7235.2010.2  and 2.1.1/4540 of the Russian Ministry for
Science and Education, and by the RFBR grants 09-02-01284-a,
11-02-00467-a. Four of us, S. T., S. Ka., E. S., and D. V.,
are grateful to the Institut f\"ur Kernphysik, Forschungszentrum
J\"ulich for its hospitality.

{}

\end{document}